\documentclass[twocolumn]{aastex63}
\usepackage{graphicx}
\usepackage{amsmath}
\usepackage{amssymb}
\usepackage{hyperref}

\def\gtrsim{\mathrel{\hbox{\rlap{\hbox{\lower4pt\hbox{$\sim$}}}\hbox{$>$}}}}
\def\lesssim{\mathrel{\hbox{\rlap{\hbox{\lower4pt\hbox{$\sim$}}}\hbox{$<$}}}}
\def\gtrsim{\mathrel{\hbox{\rlap{\hbox{\lower4pt\hbox{$\sim$}}}\hbox{$>$}}}}

\begin{document}

\title{{\sl Swift} Multiwavelength Follow-up of LVC S200224ca and the Implications for Binary Black Hole Mergers}

\correspondingauthor{N.~J. Klingler}
\email{njk5441@psu.edu}

\author[0000-0002-7465-0941]{N.~J.~Klingler}
\affiliation{Department of Astronomy and Astrophysics, The Pennsylvania State University, University Park, PA 16802, USA}
\author{A.~Lien}
\affiliation{Center for Research and Exploration in Space Science and Technology (CRESST) and
NASA Goddard Space Flight Center, Greenbelt, MD 20771, USA}
\affiliation{Department of Physics, University of Maryland, Baltimore County, 1000 Hilltop Circle, Baltimore, MD 21250, USA}
\author[0000-0001-9309-7873]{S.~R.~Oates}
\affiliation{School of Physics and Astronomy, University of Birmingham, B15 2TT, UK}
\author[0000-0002-6745-4790]{J.~A.~Kennea}
\affiliation{Department of Astronomy and Astrophysics, The Pennsylvania State University, University Park, PA 16802, USA}
\author[0000-0002-8465-3353]{P.~A.~Evans}
\affiliation{School of Physics and Astronomy, University of Leicester, University Road, Leicester, LE1 7RH, UK}
\author[0000-0002-2810-8764]{A.~Tohuvavohu}
\affiliation{Department of Astronomy and Astrophysics, University of Toronto, Toronto, ON, Canada}
\author[0000-0002-9725-2524]{B.~Zhang}
\affiliation{Department of Physics and Astronomy, University of Nevada, Las Vegas, NV 89154, USA}
\author[0000-0001-5624-2613]{K.~L.~Page}
\affiliation{School of Physics and Astronomy, University of Leicester, University Road, Leicester, LE1 7RH, UK}
\author[0000-0002-2810-8764]{S.~B.~Cenko}
\affiliation{Astrophysics Science Division, NASA Goddard Space Flight Center, Greenbelt MD, 20771 USA}
\affiliation{Joint Space-Science Institute, Computer and Space Sciences Building, University of Maryland, College Park, MD 20742, USA}
\author{S.~D.~Barthelmy}
\affiliation{Astrophysics Science Division, NASA Goddard Space Flight Center, Greenbelt MD, 20771 USA}
\author{A.~P.~Beardmore}
\affiliation{Department of Physics and Astronomy, University of Leicester, University Road, Leicester, LE1 7RH, UK}
\author{M.~G.~Bernardini}
\affiliation{INAF -- Osservatorio Astronomico di Brera, via Bianchi 46, I-23807 Merate, Italy}
\author{A.~A.~Breeveld}
\affiliation{University College London, Mullard Space Science Laboratory, Holmbury St. Mary, Dorking, RH5 6NT, U.K.}
\author[0000-0001-6272-5507]{P.~J.~Brown}
\affiliation{George P.\ and Cynthia Woods Mitchell Institute for Fundamental Physics \& Astronomy, Mitchell Physics Building, Texas A.~\& M.~University, Department of Physics and Astronomy, College Station, TX 77843, USA}
\author{D.~N.~Burrows}
\affiliation{Department of Astronomy and Astrophysics, The Pennsylvania State University, University Park, PA 16802, USA}
\author{S.~Campana}
\affiliation{INAF -- Osservatorio Astronomico di Brera, via Bianchi 46, I-23807 Merate, Italy}
\author{G.~Cusumano}
\affiliation{INAF -- IASF Palermo, via Ugo La Malfa 153, I-90146, Palermo, Italy}
\author{A.~D'A\`\i}
\affiliation{INAF -- IASF Palermo, via Ugo La Malfa 153, I-90146, Palermo, Italy}
\author{P.~D'Avanzo}
\affiliation{INAF -- Osservatorio Astronomico di Brera, via Bianchi 46, I-23807 Merate, Italy}
\author{V.~D'Elia}
\affiliation{INAF -- Osservatorio Astronomico di Roma, via Frascati 33, I-00040 Monteporzio Catone, Italy}
\affiliation{Space Science Data Center (SSDC) - Agenzia Spaziale Italiana (ASI), I-00133 Roma, Italy}
\author{M.~de~Pasquale}
\affiliation{Department of Astronomy and Space Sciences, Istanbul University, Beyz{\i}t 34119, Istanbul, Turkey}
\author{S.~W.~K.~Emery}
\affiliation{University College London, Mullard Space Science Laboratory, Holmbury St. Mary, Dorking, RH5 6NT, U.K.}
\author{J.~Garcia}
\affiliation{Cahill Center for Astronomy and Astrophysics, California Institute of Technology, Pasadena, CA 91125, USA}
\author{P.~Giommi}
\affiliation{Space Science Data Center (SSDC) - Agenzia Spaziale Italiana (ASI), I-00133 Roma, Italy}
\author{C.~Gronwall}
\affiliation{Department of Astronomy and Astrophysics, The Pennsylvania State University, University Park, PA 16802, USA}
\affiliation{Institute for Gravitation and the Cosmos, The Pennsylvania State University, University Park, PA 16802}
\author{D.~H.~Hartmann}
\affiliation{Department of Physics and Astronomy, Clemson University, Kinard Lab of Physics, USA}
\author{H.~A.~Krimm}
\affiliation{National Science Foundation, Alexandria, VA 22314, USA}
\author{N.~P.~M.~Kuin}
\affiliation{University College London, Mullard Space Science Laboratory, Holmbury St. Mary, Dorking, RH5 6NT, U.K.}
\author{D.~B.~Malesani}
\affiliation{DTU Space, National Space Institute, Technical University of Denmark, Elektrovej 327, 2800 Kongens Lyngby, Denmark}
\author{F.~E.~Marshall}
\affiliation{Astrophysics Science Division, NASA Goddard Space Flight Center, Greenbelt MD, 20771 USA}
\author{A.~Melandri}
\affiliation{INAF -- Osservatorio Astronomico di Brera, via Bianchi 46, I-23807 Merate, Italy}
\author{J.~A.~Nousek}
\affiliation{Department of Astronomy and Astrophysics, The Pennsylvania State University, University Park, PA 16802, USA}
\author{P.~T.~O'Brien}
\affiliation{Department of Physics and Astronomy, University of Leicester, University Road, Leicester, LE1 7RH, UK}
\author[0000-0002-1041-7542]{J.~P.~Osborne}
\affiliation{Department of Physics and Astronomy, University of Leicester, University Road, Leicester, LE1 7RH, UK}
\author[0000-0001-7128-0802]{D.~M.~Palmer}
\affiliation{Los Alamos National Laboratory, B244, Los Alamos, NM, 87545, USA}
\author{M.~J.~Page}
\affiliation{University College London, Mullard Space Science Laboratory, Holmbury St. Mary, Dorking, RH5 6NT, U.K.}
\author{M.~Perri}
\affiliation{Space Science Data Center (SSDC) - Agenzia Spaziale Italiana (ASI), I-00133 Roma, Italy}
\affiliation{INAF -- Osservatorio Astronomico di Roma, via Frascati 33, I-00040 Monteporzio Catone, Italy}
\author{J.~L.~Racusin}
\affiliation{Astrophysics Science Division, NASA Goddard Space Flight Center, Greenbelt MD, 20771 USA}
\author{T.~Sakamoto}
\affiliation{Department of Physics and Mathematics, Aoyama Gakuin University, Sagamihara, Kanagawa, 252-5258, Japan}
\author{B.~Sbarufatti}
\affiliation{Department of Astronomy and Astrophysics, The Pennsylvania State University, University Park, PA 16802, USA}
\author{J.~E.~Schlieder}
\affiliation{Astrophysics Science Division, NASA Goddard Space Flight Center, Greenbelt MD, 20771 USA}
\author{M.~H.~Siegel}
\affiliation{Department of Astronomy and Astrophysics, The Pennsylvania State University, University Park, PA 16802, USA}
\author{G.~Tagliaferri}
\affiliation{INAF -- Osservatorio Astronomico di Brera, via Bianchi 46, I-23807 Merate, Italy}
\author{E.~Troja}
\affiliation{Astrophysics Science Division, NASA Goddard Space Flight Center, Greenbelt MD, 20771 USA}
\affiliation{Department of Physics and Astronomy, University of Maryland, College Park, MD 20742-4111, USA}

\begin{abstract}
On 2020 February 24, during their third observing run (``O3''), the Laser Interferometer Gravitational-wave Observatory and Virgo Collaboration (LVC) detected S200224ca: a candidate gravitational wave (GW) event produced by a binary black hole (BBH) merger.
This event was one of the best-localized compact binary coalescences detected in O3 (with 50\%/90\% error regions of 13/72 deg$^2$), and so the {\sl Neil Gehrels Swift Observatory} performed rapid near-UV/X-ray follow-up observations.
{\sl Swift}-XRT and UVOT covered approximately 79.2\% and 62.4\% (respectively) of the GW error region, making S200224ca the BBH event most thoroughly followed-up in near-UV ($u$-band) and X-ray to date.
No likely EM counterparts to the GW event were found by the {\sl Swift} BAT, XRT, or UVOT, nor by other observatories.
Here we report on the results of our searches for an EM counterpart, both in the BAT data near the time of the merger, and in follow-up UVOT/XRT observations.
We also discuss the upper limits we can place on EM radiation from S200224ca, and the implications these limits have on the physics of BBH mergers.
Namely, we place a shallow upper limit on the dimensionless BH charge, $\hat{q} < 1.4 \times10^{-4}$, and an upper limit on the isotropic-equivalent energy of a blast wave $E < 4.1\times10^{51}$ erg (assuming typical GRB parameters).
\end{abstract}


\section{INTRODUCTION}

On 2015 September 12 the advanced Laser Interferometer Gravitational-wave Observatory (aLIGO) began its first observing run (``O1''; \citealt{LIGO2015}), which ran until 2016 January 19.
aLIGO consists of two interferometers located in Hanford, Washington, and Livingston, Louisiana, whose unprecedented sensitivity can detect a differential strain (over a 4 km length) of less than one ten-thousandth the charge diameter of a proton.
During O1, aLIGO made the first direct detections of gravitational wave (GW) signals of astrophysical origin: GW150914, GW151226, and LVT151012 \citep{Abbott2016a,Abbott2016b,Abbott2016c}, marking the beginning of a new era in GW astronomy.
These signals originated from the merging of binary black holes (BBHs), and thus served as an observational test of general relativity in the very strong field limit (in which no deviations from theory were seen).

After upgrades, aLIGO commenced its second observing run (O2) from 2016 November 30 to 2017 August 25.
In August, aLIGO was augmented by the addition of the Advanced Virgo detector (a 3-km-long interferometer located in Cascina, Italy; \citealt{Acernese2015}), which collectively formed the LIGO/Virgo Collaboration (LVC).
The addition of the third detector greatly enhanced LVC's localization capabilities, decreasing the size of GW error regions from hundreds to tens of square degrees in the best-case scenarios.
LVC also began promptly announcing GW triggers to some observatories under a memorandum of understanding, which allowed rapid follow-up searches.
During this run, eight more GW events were detected: seven BBH mergers and a binary neutron star (BNS) merger \citep{Abbott2019a}.

The BNS merger GW 170817 marked the first detection of an electromagnetic (EM) signal confidently associated with a GW event.
A short gamma-ray burst (sGRB), GRB 170817A, was detected by the {\sl Fermi}-GBM (Gamma-ray Burst Monitor) and {\sl INTEGRAL}-SPI-ACS (SPectrometer for Integral Anti-Coincidence Shield) coincident in time with the LVC trigger \citep{Goldstein2017,Connaughton2017,Savchenko2017,Abbott2017b}\footnote{The location of the GRB was occulted by the Earth at the orbital position of {\sl Swift}-BAT at the trigger time.}.
Follow-up searches of the GRB error region resulted in the discovery of AT2017gfo (a near-infrared/optical/ultraviolet counterpart to the BNS merger produced rapidly-cooling neutron-rich material ejected during the BNS merger) and the GRB 170817A afterglow whose nonthermal emission was seen in radio and X-rays (see, e.g., \citealt{Abbott2017,Evans2017,Troja2017,Hallinan2017}).

LVC performed part of its third observing run (O3a) from 2019 April 1 to 2019 September 30, and resumed (O3b) from 2019 November 1 to 2020 March 27\footnote{O3b was originally scheduled to proceed until 2020 April 30, but was cut short due to the COVID-19 pandemic.}.
LVC also began announcing its triggers publicly and in real-time through their web page\footnote{\url{https://gracedb.ligo.org/superevents/public/O3/}}.
O3 resulted in 56 candidate GW events, including the mergers of BBHs, BNSs, and neutron star / black holes (NSBH), as well as ``burst'' (unmodeled) triggers, and, for the first time, mergers involving objects in the mass gap range\footnote{The mass gap refers to an apparent ``gap'' in the mass spectrum of NSs and BHs, as seen in the population of X-ray binaries.  The heaviest NSs are $\lesssim 2.1M_\odot$, and the lightest BHs are $\gtrsim 5 M_\odot$.  See, e.g., \citealt{Bailyn1998,Ozel2010,Ozel2016,Abbott2019}} ($\sim3-5$ $M_\odot$).

It is generally believed that the merging of isolated BBHs do not typically produce EM radiation (see, e.g., \citealt{Kamble2013}).
However, various authors have theorized that under certain situations and/or with particular BH parameters (e.g., charged black holes, or if accreting or circumstellar material is present) BBH mergers may be able to produce detectable EM radiation (see, e.g., \citealt{Loeb2016,Perna2016,Yamazaki2016,Zhang2016,Martin2018}.).

To date, there have been two purported detections of EM transients possibly associated with BBH mergers.
\citet{Connaughton2016} reported a weak 1-s-long signal above 50 keV detected by the {\sl Fermi}-GBM 0.4 s after GW 150914 (the famous first GW signal detected, from a stellar mass BBH).
The signal's localization was consistent with that of the GW event, and had a false alarm rate (FAR) of 0.0022 (2.9$\sigma$).
The authors report the event's duration and spectra resemble a sGRB, though with a 1 keV $-$ 10 MeV luminosity of $1.8_{-1.0}^{+1.5}\times10^{49}$ erg s$^{-1}$ (assuming the distance of 410 Mpc), which is an order of magnitude dimmer than the dimmest sGRBs (excluding GW 170817A; see \citealt{Wanderman2015}). 
The astrophysical origin of this signal was, however, debated by \citet{Greiner2016}, and subsequently reaffirmed by \citet{Connaughton2018}; see also \citet{Veres2019}.
The second case of possible EM radiation from a BBH merger was seen during O3 with the detection of ZTF19abanrhr (associated with AGN J124942.3+344929, a.k.a.\ AB 15, at $z=0.438$), a candidate optical counterpart to GW 190521 (formerly designated S190521g; at an estimated  $z=0.72\pm0.29$; \citealt{Abbott2020}).
\citet{Graham2020} suggested that the flaring behavior seen in this optical transient is consistent with what one would expect from a kicked binary black hole merger in the accretion disk of an active galactic nucleus (AGN; \citealt{McKernan2019}), and ruled out other source types.
\citet{Graham2020} also predicted repeat flaring events when the kicked BBH remnant reencounters the disk on timescales (in this case) of $\sim1.6$ yr, so future follow-up will shed light on the nature of this transient. 
If the above proposed associations with BBH mergers are correct, these discoveries open up exciting new prospects for the search for EM counterparts to GW events.

Another O3 GW trigger of interest is the BBH merger S200224ca.
This was one of the best-localized BBH mergers, with 50\%/90\% error regions of 13/72 deg$^2$ (respectively).
In this paper we will discuss the follow up of this event performed by the {\sl Neil Gehrels Swift} X-ray Telescope (XRT) and Ultraviolet/Optical Telescope (UVOT), as well as a search for prompt emission near the trigger time ($T_0$) with the Burst Alert Telescope (BAT).

The paper is structured as follows.
In Section 1 we describe the {\sl Swift Observatory} and LVC trigger S200224ca.
In Section 2 we present the results of a search for prompt emission with the BAT around the time of trigger.
In Section 3 we describe {\sl Swift}'s follow-up procedure for GW events, both in general and for S200224ca specifically.
In Section 4 we provide the details of XRT and UVOT data processing.
In Section 5 we discuss the X-ray and near-UV sources found in our follow-up searches (no confirmed EM counterparts were found).
And in Section 6 we discuss the implications our results have for BBH mergers.

\subsection{The Neil Gehrels Swift Observatory}
The Neil Gehrels {\sl Swift} Observatory (\citealt{Gehrels2004}) is a multiwavelength space telescope in a low Earth orbit ($P_{\rm orbit} \approx$ 96 minutes) designed to study gamma-ray bursts (GRBs).
{\sl Swift} is equipped with three instruments.
The Burst Alert Telescope (BAT; \citealt{Barthelmy2005}) consists of a coded aperture mask covering $\sim$2 sr (or about 1/6 of the sky) and operating in the 15--350 keV range.
The X-ray Telescope (XRT; \citealt{Burrows2005}) is a focused X-ray imaging detector with a circular 23.6$'$-diameter field of view (FOV), sensitive between 0.3--10 keV.
The Ultraviolet/Optical Telescope (UVOT; \citealt{Roming2005}) covers a $17'\times17'$ FOV. 
It has six filters which span the 1600--6240 \AA\ band, and a white filter sensitive from
1600--8000 \AA, as well as UV and optical grisms for spectroscopy.

When the BAT triggers on a GRB, {\sl Swift} calculates its position to an accuracy of 1--3$'$, then autonomously and promptly slews to the burst (usually within 1-3 minutes, assuming that the burst is not located within {\sl Swift}'s regions of Sun, Earth, or Moon avoidance at trigger time).
The afterglow is then observed by the XRT and UVOT, which provide arcsecond-scale localizations.

Although initially designed to study GRBs, {\sl Swift's} ability to quickly respond to and observe targets of opportunity (TOOs) has enabled it to become a ``first responder'' to transient astronomical events of all types, including the search for EM counterparts to GW events.
{\sl Swift} has performed rapid follow-up of observations during both O2 and O3, the results of which have been reported by  \citet{Evans2016a,Evans2016b,Evans2016c,Evans2017,Evans2019,Klingler2019,Page2020,Oates2020}.

\subsection{Trigger S200224ca}

On 2020 February 24 at 22:23:34 UT, all three LVC detectors triggered on S200224ca \citep{2020GCN.27184....1L}.
The event had a low FAR of $1.605\times10^{-11}$ Hz (or roughly once every 1,974 years) and a low probability of being of terrestrial origin ($P_{\rm Terrestrial} = 3.3852\times10^{-5}$).
The candidate GW event was determined to have a very high probability of being produced by a binary black hole merger (i.e., probability $P_{\rm BBH}\approx1.0$) at an estimated distance $d=1575\pm322$ Mpc. 
Additional trigger details are listed in Table \ref{tbl-S20024ca}.
The event was very-well localized, with initial 50\% and 90\% probability areas corresponding to 17 and 69 deg$^2$, respectively (these were subsequently refined to 13 and 72 deg$^2$, \citealt{2020GCN.27262....1L}).

\begin{deluxetable}{lc}
\tablecolumns{9}
\tablecaption{LVC S200224ca Details \label{tbl-S20024ca}}
\tablewidth{0pt}
\tablehead{\colhead{Parameter} & \colhead{Value} }
\startdata
Trigger date & 2020-02-24  \\
Trigger time & 22:22:34.390 UT \\
Trigger type & Compact Binary Coalescence  \\
Detectors & H1, L1, V1; (all 3) \\
Estimated distance & $1575 \pm 322$ Mpc \\
$P_{\rm BBH}$ & 0.999966 \\
$P_{\rm MG}$ & 0.0 \\
$P_{\rm NSBH}$ & 0.0 \\
$P_{\rm BNS}$ & 0.0 \\
$P_{\rm Terrestrial}$ & $3.3852\times10^{-5}$ \\
False alarm rate & $1.605\times10^{-11}$ Hz ($1/1974$ yr) \\
\hline
{\sl Swift} Follow-up Start Time & $T_0 + 336.6$ min \\
Fields observed & 672 \\
TargetID range & $703157 - 7032252 $ \\
\enddata
\tablenotetext{}{$P_{\rm (event)}$ is the probability the event was produced by a binary black hole (BBH), mass gap object (MG), neutron star + black hole (NSBH), binary neutron star (BNS), or of terrestrial origin.  }
\end{deluxetable}

\section{Search for a Counterpart In BAT Data Near $T_0$}

We performed a search for a counterpart in the BAT data within $T_0 \pm 100$ s.
Specifically, we searched for potential detections in the BAT raw light curves and images created by the BAT survey data and the available BAT event data. 
During O3, {\sl Swift} had begun testing the BAT Gamma-ray Urgent Archiver for Novel Opportunities (GUANO) system \citep{Tohuvavohu2020}.
GUANO is a system which proactively downlinks BAT event data from the spacecraft (which is otherwise discarded in the absence of a BAT trigger due to telemetry limitations) upon receiving notice of a relevant astronomical event detected by other facilities. 
The retrieval of BAT event data allows for much more sensitive searches for prompt (sGRB-like) emission from compact binary coalescence (CBC) events, and increases the number of possible co-detections of GW and EM radiation.
Unfortunately, for S200224ca, only event data from $T_0+29.042$ s to $T_0+45.176$ s were retrieved from BAT's ring buffer due to the GUANO command reaching the {\sl Swift} spacecraft a little too late\footnote{This was due to a long request queue in the Tracking and Data Relay Satellite System (TDRSS) and the unavailability of ground station passes shortly after $T_0$.}, at $\sim T_0+28$ minutes instead of the median $\sim T_0+16$ minutes for GW triggers.

Our analysis of event data and survey data utilized the standard HEASoft/BAT tools. 
Specifically, {\tt batgrbproduct} was used for the event data analysis, {\tt batsurvey} was used for survey data analysis, and {\tt batcelldetect} was used to search for potential detections in images.
Analysis of the BAT data were also reported via GCN circular by \citet{Barthelmy2020}.

\begin{figure*}
\includegraphics[width=1.0\hsize,angle=0]{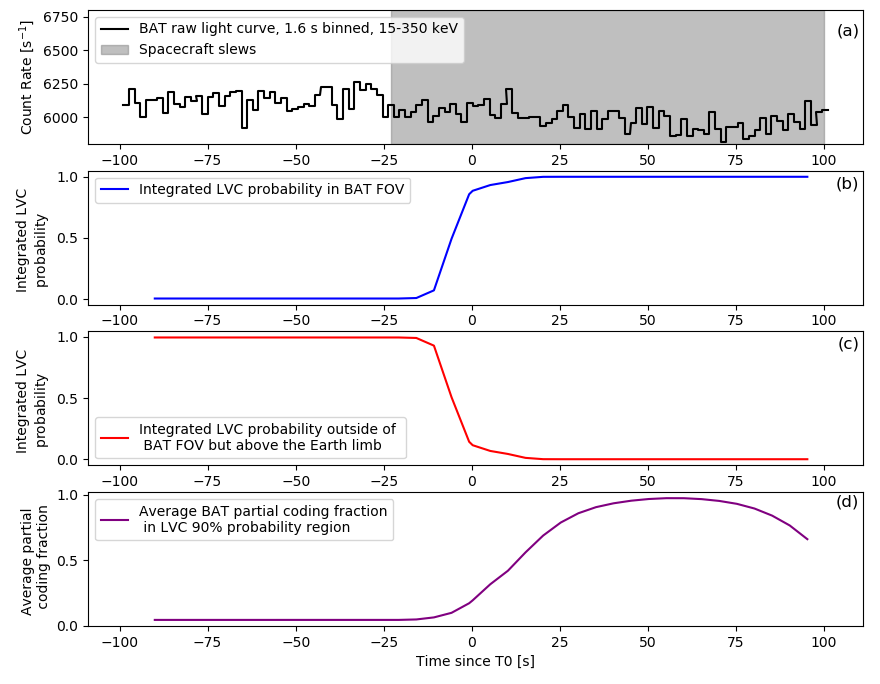}
\caption{{\sl Panel (a)}: BAT raw light curve ($15-350$ keV) consisting of 1.6-s bins. The gray area marks the time interval of spacecraft slewing.
{\sl Panel (b)}: The integrated LVC localization probability inside the BAT FOV ($> 10\%$ partial coding fraction).
{\sl Panel (c)}: The integrated LVC localization probability outside of the BAT FOV but above the Earth limb.
{\sl Panel (d)}: the average BAT partial coding fraction inside the 90\% LVC localization probability region.  The integrated LVC probability and average partial coding fraction plots correspond to the LALInference map (i.e., the most recent sky map for this event).}
\label{fig-BAT-FOV}
\end{figure*}

{\sl Swift} was slewing from $\sim T_0-23$ s to $\sim T_0+180 $ s. 
Figure~\ref{fig-BAT-FOV} shows the BAT raw light curve (panel a) and how the BAT FOV coverage of the LVC probability region changes as a function of time.
As BAT is a coded aperture mask instrument, the sensitivity is highest at the center of the FOV, where the coding fraction is $\sim 100\%$. 
The sensitivity decreases toward the edge of the FOV, as the coding fraction decreases to zero\footnote{For more details on the coded aperture mask instrument and the definition of the coding fraction, see the {\sl Swift} BAT Software Guide, \url{http://swift.gsfc.nasa.gov/analysis/bat_swguide_v6_3.pdf}}. 
The integrated LVC localization probability inside the BAT FOV ($> 10\%$ partial coding fraction) is shown in panel (b). 
In addition, BAT retains decreased (but still significant) sensitivity to rate increases for gamma-ray events outside of its FOV. 
In panel (c) we also show the integrated LVC localization probability that is outside of the BAT FOV, but above the Earth limb. 
At $T_0$, the S200224ca integrated probability inside the BAT FOV is 0.8855, and the probability outside of BAT FOV but above the Earth limb is 0.1145.
The BAT FOV at $T_0$ and S200224ca localization are also overlayed in the right panel of Figure \ref{fig-xrt-convolved}.

Within our search time window of $T_0 \pm 100$ s, the BAT raw light curves in different time bins (64 ms, 1 s, 1.6 s) and energy bands ($15-25$, $25-50$, $50-100$, and $100-350$ keV) show no significant detections with signal-to-noise ratio $\gtrsim 5\sigma$. 
We estimate the flux upper limit using the 1.6-s light curve ($15-350$ keV). 
Within $T_0 \pm 100$ s, the light curve has a $5 \sigma$ standard deviation of $\sim 411$ count/s, which corresponds to a $5 \sigma$ flux upper limit of $\sim 4.76 \times 10^{-7}$ erg cm$^{-2}$ s$^{-1}$ (in $15-350$ keV). 
The conversion from instrumental count to energy flux uses the method developed by \citet{Lien2014}, and assumes a typical BAT-detected short GRB spectrum (a simple power-law model with a photon index of $1.32$; \citealt{Lien2016}) and an instrumental response for a partial coding fraction of $\sim 0.19$, which is the average partial coding fraction of the LVC region inside the BAT FOV at $T_0$. 

The event data closest to $T_0$ only cover the time range of $T_0+29.04$ s to $T_0+45.18$ s.
No significant detections ($\gtrsim 5\sigma$) are found in the image ($15-150$ keV) made using these event data. 
Because the time interval covers a spacecraft slew, this image is made from co-adding sub-images of 0.5-s intervals. 

BAT collects continuous survey data (i.e., data binned in intervals of $\sim 300$ s) while the spacecraft is in pointing mode. 
Because of the spacecraft slews, no survey data are available at $T_0$. 
The closest available survey data cover time ranges of $T_0-144.96$ s to $T_0-21.96$ s, and $T_0+180.02$ s to $T_0+480.04$ s. 
Again, no significant detections are found in images created by survey data in these two time intervals.

\section{Swift GW Follow-Up}
{\sl Swift's} criteria and procedure for performing pointed follow-up searches of GW error regions (with UVOT/XRT) have been described in detail by \citet{Evans2016a,Evans2016c}, and subsequent improvements have been described by \citet{Klingler2019} and \citet{Page2020}, so only an abridged description will be provided here, in Section 3.1.
In Section 3.2 we describe the observing plan for S200224ca.

\subsection{General Follow-up Procedure}
Upon notification of a new LVC trigger, our GW planning pipeline determines whether our follow-up criteria are met.
Factors which determine this include the type of event (e.g., CBC or burst event), the probability that the GW event involved a neutron star $P_{\rm NS}$, the event's estimated FAR, the probability the signal was of terrestrial origin $P_{\rm Terrestrial}$, and the percentage of the probability region that {\sl Swift}-XRT can observe within 24 hours $P_{\rm 24hrs}$ (the full follow-up criteria are described by \citealt{Page2020}).

When the 3D LVC probability map is received, it is automatically convolved with the 2MASS Photometric Redshift Galaxy Catalog (2MPZ; \citealt{Bilicki2014}), using the method described by \citet{Evans2016c,Evans2019}.
This approach accounts for galaxy catalog (in)completeness. 
For events at large distances where galaxy catalogs are highly incomplete, as is the case for 200224ca, the convolved sky map is only minimally altered from the original LVC map.
We use this reweighted probability map to create a prioritized observing plan.

\subsection{S200224ca Follow-up}

\begin{figure*}
\includegraphics[width=0.98\hsize,angle=0]{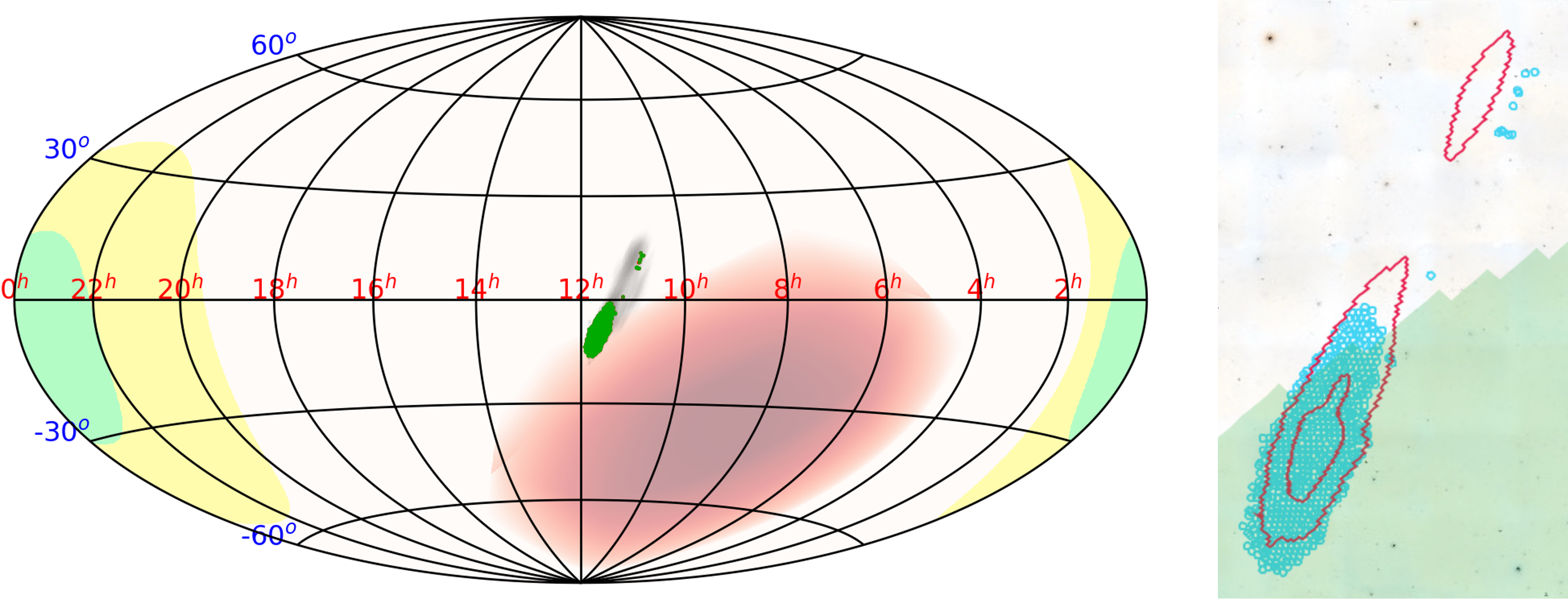}
\caption{{\sl Left:} Updated skymap of the S200224ca localization (from the LALInference pipeline; shown by the gray area).  The area covered by the XRT follow-up is highlighted in dark green, and the region of sky covered by BAT's coded aperture mask at the time of trigger ($T_0$) is highlighted in red.  The yellow and light green areas represent {\sl Swift's} areas of Sun and Moon avoidance at $T_0$.  {\sl Right:} Zoomed-in image of the central regions of the S200224ca localization, overlayed on a DSS image (produced by \url{treasuremap.space}; \citealt{Wyatt2020}).  The 50\%/90\% contours are enclosed in red, the XRT fields are circled in blue, and the region of sky covered by BAT (i.e., $>10\%$ coding fraction) at $T_0$ is highlighted in green.  The XRT fields outside of the northern (50\%) lobe were due to {\sl Swift} performing observations prior to the release of the updated (LALInference) skymap.}
\label{fig-xrt-convolved}
\end{figure*}

\begin{figure}
\includegraphics[width=0.98\hsize,angle=0]{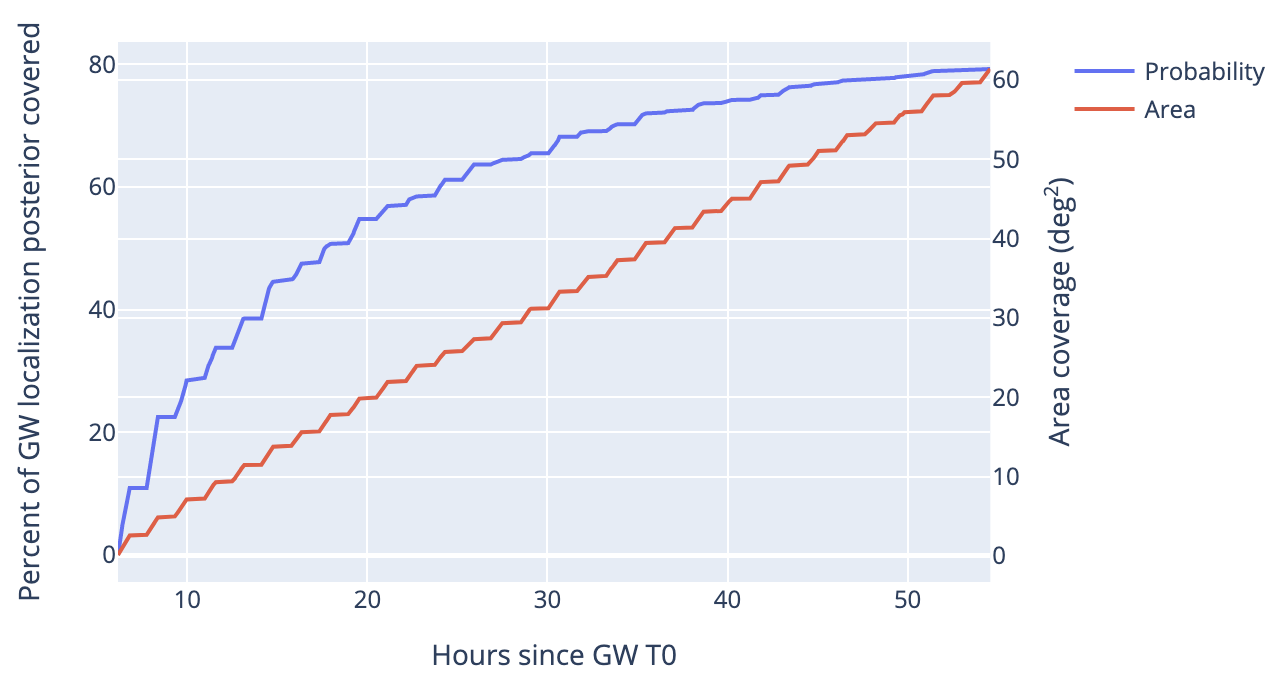}
\caption{Percentage of the S200224ca GW localization (blue) and raw area (red) covered by the XRT as a function of time (using the latest LALInference skymap; from \url{treasuremap.space}; \citealt{Wyatt2020}). The step-like behavior of the plot results from a combination of {\sl Swift}'s low-Earth orbit and Earth-limb pointing constraints, which causes any given region of sky to be only visible for (at most) roughly half of the $\approx$96-minute orbit.}
\label{fig-XRT-coverage}
\end{figure}

\begin{figure}
\includegraphics[width=0.98\hsize,angle=0]{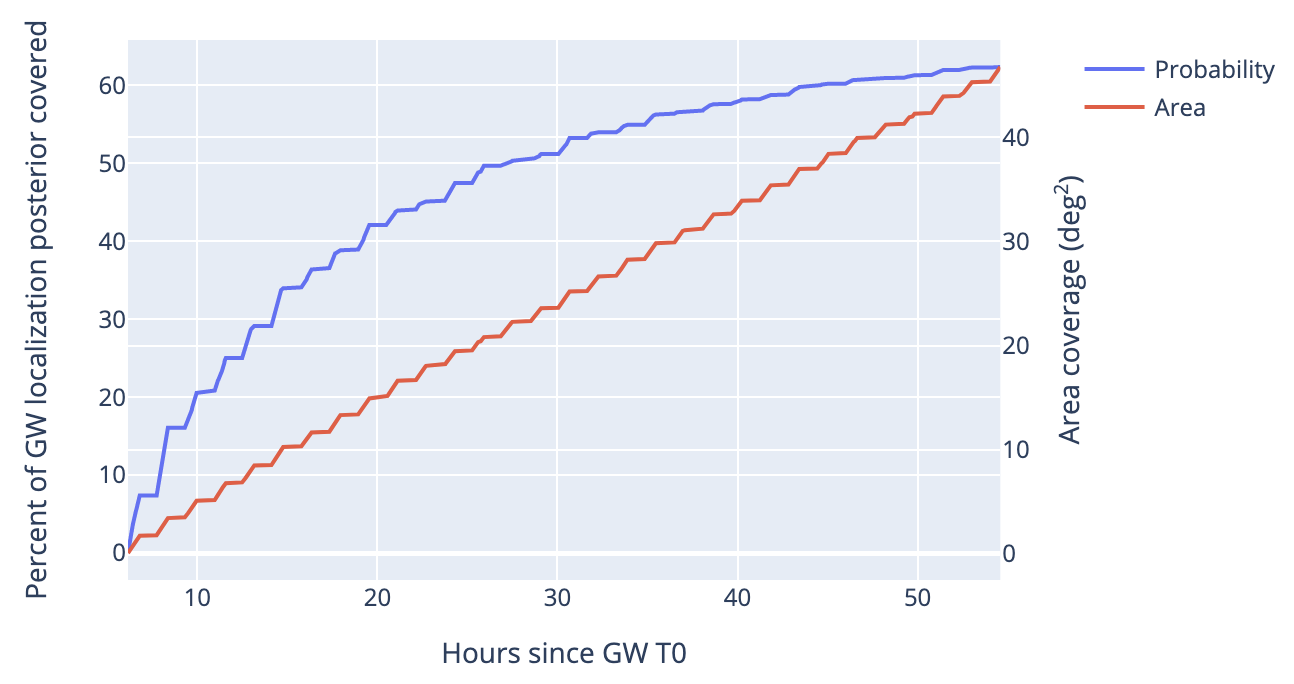}
\caption{Percentage of the S200224ca GW localization (blue) and raw area (red) covered by the UVOT as a function of time (using the latest LALInference skymap; from \url{treasuremap.space}; \citealt{Wyatt2020}). }
\label{fig-UVOT-coverage}
\end{figure}

Upon receiving automated notice of the trigger, our LVC processing pipeline calculated $P_{\rm 24hrs}=0.66$, which met our criteria for follow-up.
A tiling plan was created and uploaded to {\sl Swift} at the next available ground station pass, and {\sl Swift} began observations at $T_0+22$ ks. 
672 fields were observed (the UVOT filter wheel was in ``blocked'' mode for 2 of these fields due to the presence of bright stars).
{\sl Swift} completed the phase 1 (80 s) tilings at $T_0+196.5$ ks, but the planned phase 2 (500 s) tilings were not carried out due to the announcement of the following GW event, S200225q, whose follow-up took priority.
The XRT covered 61.2 deg$^2$ (taking into account tiles with overlapping regions), corresponding to 79.2\% of the GW probability region (using the most recent LALInference skymap\footnote{See \url{https://gracedb.ligo.org/superevents/S200224ca/}}).
In Figure \ref{fig-xrt-convolved} we show the S200224ca localization with the area covered by the XRT overlayed, and in Figure \ref{fig-XRT-coverage} we show the percentage of the S200224ca covered by the XRT as a function of time.
The UVOT performed follow-up in the $u$-band (near-UV) and covered 46.2 deg$^2$ (due to its FOV being smaller than the XRT's), corresponding to 62.4\% of the raw probability region (see Figure \ref{fig-UVOT-coverage}).

\section{DATA PROCESSING}

\subsection{XRT Data Processing}
Upon being downlinked, XRT data were automatically processed at the United Kingdom {\sl Swift} Data Science Centre (UKSSDC) at the University of Leicester, using HEASoft v6.26.1 and the latest {\tt CALDB} available at the time of processing.
The {\tt xrtpipeline} tool is run, which applies all necessary calibrations, filtering, and corrections\footnote{For more details, see \url{http://www.swift.ac.uk/analysis/} and \url{https://heasarc.gsfc.nasa.gov/ftools/caldb/help/xrtpipeline.html}}.
Images and exposure maps of each observation are also created.

The goals of the subsequent data processing are as follows: (1) search for sources, (2) characterize them, (3) identify potential counterparts to the GW trigger.

The source detection pipeline is that used in the production of the 2nd {\sl Swift} X-ray Point Source Catalog (2SXPS), which is described in detail by \citet{Evans2020}.
This is an iterative process which utilizes sliding-cell source detection, background modeling, PSF fitting, and a likelihood test to detect and localize sources.
Each detected source is assigned a quality flag, which describes the probability of the source being a spurious detection.
Sources flagged as ``good'' have a 0.3\% or less chance of being spurious (or false positive; FP), ``reasonable'' sources have a 7\% FP rate, and ``poor'' sources have up to a 35\% FP rate.
All sources detected are manually inspected for spurious detections that may arise from optical loading, stray light, extended emission, and/or thermal noise from the XRT detector getting too hot.

For source characterization, the sources are assigned a rank which indicates how likely each is to be an EM counterpart to a GW trigger, as defined by \citet{Evans2016b} (this is same ranking system used during {\sl Swift} follow-up of O2 triggers; see \citealt{Klingler2019}).
Rank 1 is assigned to counterpart candidates.
Sources of this rank are: (1) uncataloged and at least $5\sigma$ above the $3\sigma$ upper limit from the {\sl ROSAT} All-Sky Survey (RASS) or 2SXPS, or (2) a known X-ray source which is $5\sigma$ above its cataloged flux\footnote{The archival upper limits / count rates (i.e., from RASS, or 2SXPS when available) for both criteria were not derived from XRT data; they have been converted to equivalent XRT (PC mode) 0.3--10 keV count rates using {\tt PIMMS} (Portable Interactive Multi-Mission Simulator), assuming a typical AGN spectrum (absorbing Hydrogen column density $N_{\rm H}=3\times10^{20}$ cm$^{-2}$, and photon index $\Gamma=1.7$).  The peak source fluxes were also obtained by converting from the peak count rates when assuming a typical AGN spectrum with the above-mentioned parameters.}.
Rank 1 sources must also lie near (within 200 kpc in projection of) a known galaxy (assuming the source is at the distance of that galaxy).
Rank 2 is assigned to ``interesting'' sources.
These are either: (1) uncataloged and: at least $3\sigma$ above the $3\sigma$ upper limit from the RASS/2SXPS or are fading, or (2) known X-ray sources at least 3$\sigma$ above their catalogued flux.
Unlike afterglow candidates, an interesting source need not be near a known galaxy.
Rank 3 is assigned to X-ray sources not previously cataloged, and which were not bright enough to suggest a transient nature.
Rank 4 is assigned to known X-ray sources, and whose flux is consistent with (or below) that from their previous observations.

Light curves are produced for each detected source using all follow-up data, and sources exhibiting evidence of fading are flagged.

\subsection{UVOT Data Processing}
Analysis of the UVOT data is performed via the UVOT GW processing pipeline, for which we provide a brief overview here (see \citealt{Oates2020} for full details). 
The pipeline searches the UVOT sky images to identify new transient sources that might be the counterpart to the GW event. 
For each observation, the HEASoft {\tt ftool} utility {\tt uvotdetect} (based on {\tt SExtractor}; \citealt{Bertin1996}) is run for that exposure to search for sources. 
All sources found are then run through a series of checks to determine if they are previously cataloged (known) sources, extended sources, sources due to image artifacts, or minor planets.
Sources that pass all quality-control checks are labelled as Q0 or Q1, depending on their magnitude.
Sources dimmer than a magnitude of 18.9 (the typical sensitivity limit for tiling observations) are assigned as Q1. 
The pipeline reliably finds new sources if they are isolated from existing sources and not affected by defects in the UVOT images (artifacts) caused by very bright sources. 
Small images (thumbnails) are produced for each Q0 and Q1 source, and also for all nearby galaxies reported in the GLADE Catalog \citep{Dalya2018} whose positions were observed by UVOT. 
This enables rapid manual inspection to evaluate the reliability of possible UVOT counterparts. 
Also, for every Q0/Q1 UVOT source found, the XRT analysis is automatically repeated to determine flux measurements at the UVOT source position, which is more sensitive than a blind search. 
If the 3$\sigma$ lower-limit on the XRT count rate is $>$0, then an X-ray counterpart to the UVOT source is reported as detected; otherwise the 3-sigma UL is reported. 
In the case of S200224ca, only upper limits were found.

\section{S200224ca Follow-up Results}

No confirmed or likely counterparts to S200224ca were detected by {\sl Swift}, or by other observatories.
Below we describe the (uncataloged) XRT sources and the near-UV sources detected and flagged by the XRT/UVOT  pipelines.
XRT count rates provided were corrected for vignetting effects.

\subsection{X-ray Sources}

{\sl Swift}-XRT detected 8 sources during the follow-up of S200224ca. 
Two were rank 3 (uncataloged X-ray sources, though not significantly above the RASS upper limit), and six were rank 4 (known X-ray sources whose fluxes were consistent within 3$\sigma$ of their cataloged values).
Details for each source are listed in Table \ref{tab-sources}.
Source 9 had a detection flag of ``reasonable'', and the rest had detection flags of ``good''.

\begin{deluxetable*}{lcccccccccccc}
\tablecolumns{13}
\tablecaption{Detected XRT Sources}
\tablehead{
\colhead{$\#$} & \colhead{Rank} & \colhead{R.A./Dec.} & \colhead{Err.} & \colhead{Exposure} & \colhead{Peak} & \colhead{Peak} & \colhead{Fading} &  \colhead{$D_{\rm LVC}$} &  \colhead{Peak} &   \colhead{$F_{\rm Obs}$:$F_{\rm Cat}$} &   \colhead{G/2/S} & \colhead{Gal.} \\ 
\colhead{} & \colhead{} & \colhead{(J2000)} & \colhead{(90\%)} & \colhead{} & \colhead{Rate} & \colhead{$F_{-12}$} & \colhead{} &  \colhead{} &  \colhead{$L_{44}$} &   \colhead{} &   \colhead{} & \colhead{Comp.} \\
\colhead{} & \colhead{} & \colhead{} & \colhead{arcsec} & \colhead{s} & \colhead{ct s$^{-1}$} & \colhead{(cgs)} & \colhead{$\sigma$} &  \colhead{Mpc} &  \colhead{erg s$^{-1}$} &   \colhead{$\sigma$} &   \colhead{} & \colhead{}
}
\startdata
5 & 3 & 173.8795 -12.7028 & 6.3 & 1880 & $0.09^{+0.05}_{-0.04}$ & $3.7^{+2.1}_{-1.6}$ & 0 & $1391\pm363$ & $8.5\pm1.8$ & 0 & 0/0/0 & 4.7\% \\
9 & 3 & 174.3296 -4.7332 & 6.2 & 7032 & $0.09^{+0.05}_{-0.04}$ & $3.7^{+2.1}_{-1.6}$ & 2.3 & $1428\pm298$ & $9.1\pm2.0$ & 2.9 & 0/0/0 & 3.6\% \\
\hline
1 & 4 & 175.5795 -14.3774 & 5.2 & 80 & $0.09^{+0.05}_{-0.03}$ & $3.9^{+2.2}_{-1.3}$ & ... & $1551\pm334$ & $11\pm2.2$ & 1.4 & 0/2/1 & 3.7\% \\
3 & 4 & 173.9797 -11.7069 & 5.9 & 65 & $0.07^{+0.05}_{-0.03}$ & $2.9^{+2.1}_{-1.2}$ & ... & $1470\pm335$ & $7.5\pm1.8$ & 0 & 0/2/1 & 3.9\% \\ 
4 & 4 & 162.5324 +11.5419 & 4.5 & 95 & $0.21^{+0.08}_{-0.07}$ & $8.9^{+3.4}_{-3.0}$ & 1.4 & $1587\pm347$ & $3.7\pm0.6$ & 2.8 & 0/1/1 & 12.5\% \\ 
7 & 4 & 175.4238 -14.1306 & 5.2 & 137 & $0.09^{+0.05}_{-0.04}$ & $3.8^{+2.1}_{-1.7}$ & 0 & $1544\pm332$ & $1.1\pm0.2$ & 0 & 0/1/1 & 3.7\% \\
8 & 4 & 171.4663 -7.7071 & 5.1 & 313 & $0.06^{+0.02}_{-0.02}$ & $2.5^{+0.8}_{-0.6}$ & ... & $1173\pm394$ & $4.1\pm0.5$ & 2.8 & 0/1/1 & 6.3\% \\
10 & 4 & 178.0148 -11.3727 & 4.7 & 155 & $0.10^{+0.05}_{-0.04}$ & $4.3^{+2.2}_{-1.7}$ & 0 & $1432\pm394$ & $10.4\pm2.0$ & 0 & 0/1/1 & 5.0\% 
\enddata
\label{tab-sources}
\tablenotetext{}{Details of the sources detected.  The following are listed: source number (sources were numbered in the order of detection; missing numbers were sources confirmed to be spurious detections), source rank, position and 90\% uncertainty, total XRT exposure (corrected for vignetting effects), peak count rate (0.3--10 keV), peak flux $F_{-12}$ (in units of $10^{-12}$ erg cm$^{-2}$ s$^{-1}$), the significance of any fading (``...'' notes sources observed once), the LVC estimated distance at the source's position $D_{\rm LVC}$, and the peak X-ray luminosity $L_{44}$ (if the source was located at $D_{\rm LVC}$, in units of $10^{44}$ erg s$^{-1}$).  $F_{\rm Obs}$:$F_{\rm Cat}$ shows the significance of the ratio of the observed flux to the cataloged flux; for uncataloged (rank 3) sources, the XRT-equivalent of the RASS upper limit was used instead.  
The G/2/S column notes whether the source's position (and uncertainty) are consistent with any known GWGC/2MPZ galaxies, 2MASS sources, or SIMBAD sources.  Gal.\ Comp.\ is the completeness of the galaxy catalog along the line of sight to the source.}
\end{deluxetable*}

\begin{figure}
\includegraphics[width=0.98\hsize,angle=0]{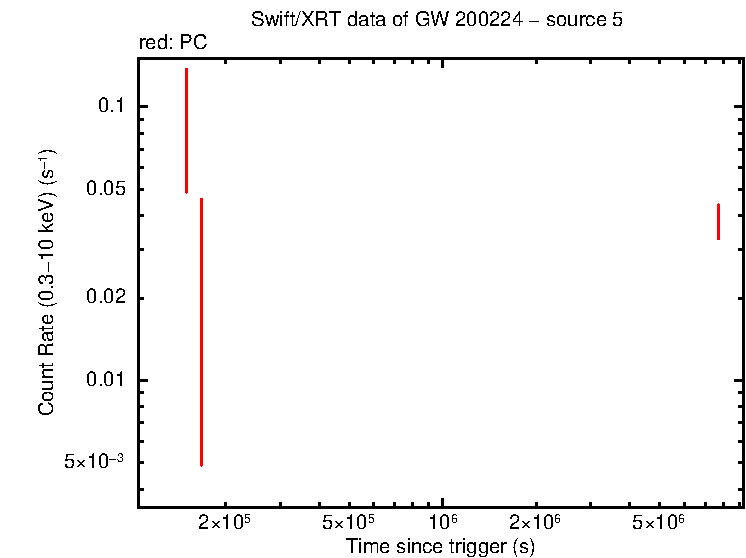}
\caption{XRT light curve of Source 5.}
\label{fig-source-5}
\end{figure}

The two rank 3 sources were Sources 5 and 9.
Source 5 was first detected during Observation 07032074001, which began on 2020-02-26 at 15:56:43 UT.
During the 82 s exposure the source was observed to have a count rate of $0.09^{+0.05}_{-0.04}$ ct s$^{-1}$.
It was observed again during Observation 07032138001, which began at 2020-02-26 at 20:38:59 UT, during which the count rate was $0.02^{+0.03}_{-0.01}$, thus exhibiting evidence of fading at $1.3\sigma$.
Source 5 was reobserved for 1.7 ks on 2020 May 24, after which it was found to have shown no evidence of fading after 6 months (see the light curve in Figure \ref{fig-source-5}).
The nearest cataloged source is quasar J113530.8-124219, located 10.6$''$ from the XRT position (though it is outside of the 90\% positional uncertainty radius of 6.3$''$.

\begin{figure*}
\includegraphics[width=0.98\hsize,angle=0]{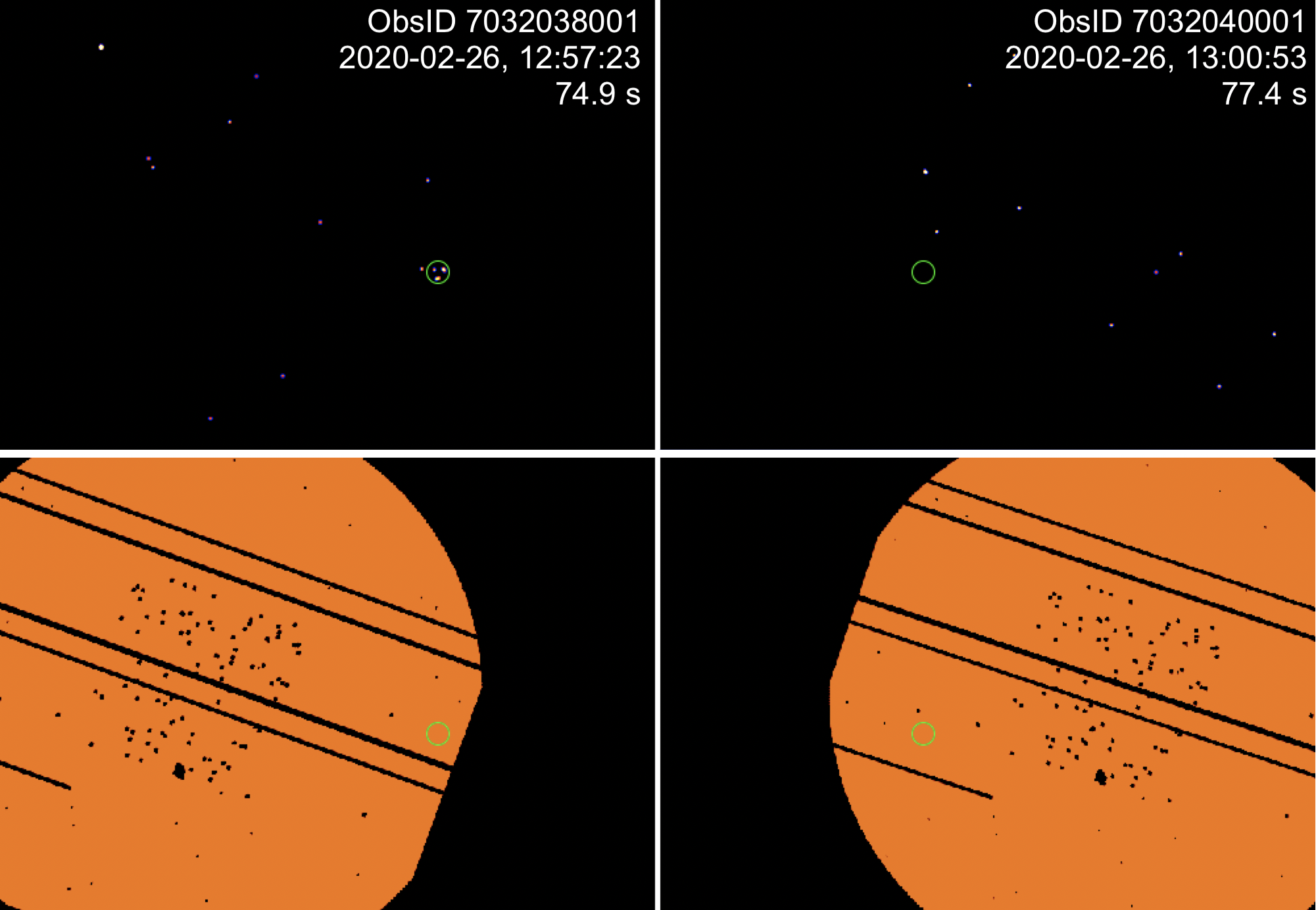}
\caption{{\sl Top:} XRT images from the first two observations of the field containing Source 9. {\sl Bottom:} Exposure maps of the above observations.  The exposure maps show where the source lies in the XRT FOV, and the locations of dead pixels and hot columns (which are masked out).}
\label{fig-source-9}
\end{figure*}

Source 9 was detected during Observation 07032038001, which began on 2020-02-26 at 12:57:23 UT ($\approx T_0 + 139$ ks).
During the 75 s exposure, 5 counts were detected (within a 30$''$ aperture radius), yielding a count rate of $0.09^{+0.05}_{-0.04}$ ct s$^{-1}$.
The target's location was reobserved a few minutes later at 13:00:53 (since the target resided in the overlapping region between two planned pointings; the second of which was Observation 07032040001), but no counts were detected during the 77 s exposure.
The XRT images and exposure maps of Observations 07032038001 and 07032040001 are shown in Figure \ref{fig-source-9}.
In the first observation (07032038001), Source 9's position did not land within the UVOT FOV.
In the second observation (07032040001), Source 9's position resided in the UVOT's FOV (at the very edge), but no near-UV counterpart was seen.

Source 9's position was reobserved four times between May and June 2020 (totaling 8.2 ks; targetID 7400091) but no counts were detected within a 30$''$ aperture, yielding a count rate $<1.2\times10^{-4}$ ct s$^{-1}$.
An 8.2 ks exposure corresponds to a detection limit of approximately $7.5\times10^{-14}$ erg cm$^{-2}$ s$^{-1}$ at 50\% confidence (or $1.5\times10^{-13}$ erg cm$^{-2}$ s$^{-1}$ at 90\% confidence; see the XRT sensitivity curve in Figure 6 of \citealt{Evans2015}).
No near-UV counterpart was seen in any of the corresponding UVOT images.
There are no cataloged SIMBAD sources within 18$''$ of this source.
The nearest VizieR cataloged source is SDSS J113719.13-044407.7, located 8$''$ from the XRT position (though it is outside the 90\% positional uncertainty radius of 6.2$''$).

The detection of Souce 9 in ObsID 7032038001 but not in ObsID 7032040001 (which occurred 4 minutes later) is puzzling.
Assuming that Source 9's count rate is Poissonian (and that it did not change noticeably over the course of the 4 minutes), the probability of detecting 0 counts during the second (77.4 s) exposure is low: 0.000943.
Considering this, it is possible that the detection of Source 9 in the first observation was a false positive resulting from noise fluctuations, as its detection flag was only ``reasonable'' (sources of this flag have a 7\% false positive rate).
Even if Source 9 was real, such fading over the course of minutes at $T_0 + 139$ ks is not consistent with power-law fading observed in sGRBs (see, e.g., \citealt{Fong2015}; although BBH mergers are not expected to produce sGRBs under typical circumstances).

\subsection{Near-UV Sources}
The UVOT GW processing pipeline produced 1654 thumbnails from the 670 observed fields. 
The majority of these thumbnails were of galaxies, but 17 and 65 thumbnails were produced for Q0 and Q1 sources, respectively. 
Of these, manual inspection ruled out the majority of these sources, but 1 Q0 source and 2 Q1 sources were still of interest after manual inspection.
Additional discussion of these sources and a table listing the photometry is given in \citet{Oates2020}.

Q0\_src47 was found at RA, Dec (J2000) = 176.58794, $-$11.46508 with an estimated uncertainty of 0.8$''$ (radius, 90\% confidence), starting at $T_0 + 63.77$ ks. 
The initial $u$-band magnitude\footnote{The UVOT source magnitudes listed in this work correspond to the Vega magnitude system.  Absolute magnitudes (AB) are provided in parentheses.} was $u = 18.62\pm0.23$ ($19.64$ mag.\ AB), giving a $4.8\sigma$ detection. 
This source was announced to the astronomical community via \citet{Breeveld2020}. 
No follow-up was reported by other facilities. 
In the Pan-STARRS catalog a point-like source was found (objID 94241765879572233) at a position consistent with this Q0 source with (AB) magnitudes
$z=20.74\pm0.04$, $i=21.59\pm0.23$.
Pan-STARRS did not detect the source in the $g$- or $r$-bands, with limiting magnitudes (5$\sigma$, AB) $>23.3$ and $>23.2$, respectively.
Q0\_src47 was not detected in any filter in the second UVOT visit ($u>19.6$) occurring at occurring at $T_0 + 3.18$ days (or 2.5 days after the first detection); thus, the target faded by $\Delta u > 0.4$ mag day$^{-1}$.
The observed rapid fading rules out slowly-evolving transients such as such as SNe \citep{Wheeler1990,GalYam2012}, tidal disruption events (TDEs; \citealt{vanVelzen2020}), and AGN \citep{MacLeod2012,Smith2018}, but faster evolving transients such as GRBs \citep{Sari1998}, kilonovae \citep{Metzger2019} and flare stars \citep{Osten2010,Schmidt2014} can not be excluded. 
Archival observations by PanSTARRS shows that Q0\_src47 is point-like and very red. 
Its brightness increases toward red wavelengths and it is at least two magnitudes fainter in $g$ and $r$ compared to $i$, $z$ and $Y$. 
Considering the target's point-like nature, red quiescent color, and rapid fading, it is likely that what UVOT detected was a flare from a low-mass star (see also \citealt{Oates2020} for further analysis of this source). 
This source was reobserved again on 2020-06-30 ($T_0 + 127$ days), but was not detected ($u>20.8$).

Q1\_src40 was found at RA, Dec (J2000) = 173.27717, $-$2.44852 with an estimated uncertainty of 0.8$''$ (radius, 90\% confidence), starting at $T_0 + 2.13$ days. 
The initial $u$-band image gave a 3.2$\sigma$ detection with a magnitude of $19.42\pm0.34$  ($20.44$ mag.\ AB). 
A second longer UVOT image (258s) gave a 5.5$\sigma$ detection with a $u$-band magnitude of $19.54\pm0.20$ ($20.56$ mag.\ AB) taken at $T_0 + 10.54$ days. 
A cataloged source, SDSS J113306.49--022654.8, is consistent with the location of this source, and with archival magnitude $u=21.168\pm0.081$ (AB).
This source is listed in several catalogs as a candidate AGN/quasar \citep{Suchkov2005,Brescia2015,Nakoneczny2019}.
The observed brightening (compared to the archival magnitude) is likely due to AGN variability. 
However, if the BBH merger occurred in an AGN accretion disk, the kicked remnant BH could disturb the disk making AGN brighten (see, e.g., \citealt{Graham2020}), although such a scenario can not be confidently discerned with the available data.

Q1\_src54 was identified at a RA, Dec (J2000) = 173.30543, $-$2.46977 with an estimated uncertainty of 0.8$''$ (radius, 90\% confidence), starting at $T_0 + 2.13$ days. 
The 75-s image gave a 3.4$\sigma$ detection with a $u$-band magnitude of $19.04\pm0.32$ ($20.06$ mag.\ AB). 
The source was not detected in a second longer UVOT $u$-band image (258s) with a 3$\sigma$ upper limit of $> 19.94$ taken at $T_0 + 10.54$ days, suggesting the source has faded between observations. 
A deeper (905 s) $u$-band observation taken at $T_0 + 136$ days failed to find any source, down to a magnitude of $>20.44$ ($>21.46$ AB).
Q1\_src54 was found to have a WISE counterpart whose position is consistent within 4$''$, which may suggest an underlying red progenitor, potentially associating this $u$-band detection as a possible stellar flare from a cool star (see \citealt{Oates2020} for details).

\section{Physical implications}

No credible or likely EM counterpart to S200224ca was detected in our follow-up XRT/UVOT observations, in the BAT data around the time of trigger, or by other observatories.
The lack of detection of any candidate counterparts, though, is a confirmation of the expected (null) result, or that emission (if present) was below the sensitivity threshold for the UVOT/XRT/BAT, as well as for those of the other observatories which followed-up this event.

With the trigger distance, $d=1575\pm322$ Mpc, the $\sim 4.76 \times 10^{-7}$ erg cm$^{-2}$ s$^{-1}$ BAT upper limit (in the $15-350$ keV band) corresponds to a gamma-ray luminosity upper limit of $1.4^{+0.6}_{-0.5} \times 10^{50} \ {\rm erg \ s^{-1}}$. 
In 80 s exposures, the XRT reaches a sensitivity of $\approx 4.2\times10^{-12}$ erg cm$^{-2}$ s$^{-1}$ at 50\% confidence, and $\approx10^{-11}$ erg cm$^{-2}$ s$^{-1}$ at 90\% confidence (0.3--10 keV).
These correspond to luminosity upper limits of $1.2_{-0.4}^{+0.6}\times10^{45}$ and $3.0_{-1.1}^{+1.3}\times10^{45}$ erg s$^{-1}$, respectively, at the trigger distance.
With the same exposures, the UVOT can reach an approximate $u$-band limiting magnitude of $19.2$ (Vega), or $20.2$ (AB).
This corresponds to a flux density of $<7.04\times10^{-17}$ erg cm$^{-2}$ s$^{-1}$ {\AA}$^{-1}$ (at 3465 {\AA}, the center of UVOT's $u$-filter), and a luminosity ($\nu L_\nu$) upper limit of $7.2_{-2.7}^{+3.3}\times10^{43}$ erg s$^{-1}$ (at $8.65\times10^{14}$ Hz).
The uncertainties on the luminosity upper limits correspond to the uncertainty in estimated merger distance.

It is worth noting that UVOT's upper limits are not as constraining as those provided by larger ground-based optical observatories that covered the event, although UVOT's limits are at bluer wavelengths.
For example, Subaru-HSC covered 83.7\% of the GW error region between $T_0+12.3$ hr and $T_0+13.4$ hr, and obtained limiting (AB) magnitudes in the $r$- and $z$-band of $>24.8$ and $>23.5$, respectively \citep{Ohgami2020}.
These Subaru limits correspond to flux densities of $<3.38\times10^{-19}$ erg cm$^{-2}$ s$^{-1}$ {\AA}$^{-1}$ (at 6222 {\AA}) and $<5.43\times10^{-19}$ erg cm$^{-2}$ s$^{-1}$ {\AA}$^{-1}$ (at 8917 {\AA}).
These tighter flux density upper limits place luminosity ($\nu L_\nu$) upper limits $6.2^{+2.7}_{-2.2}\times10^{41}$ erg s$^{-1}$ at $4.82\times10^{14}$ Hz, and $1.43^{+0.65}_{-0.52}\times10^{42}$ erg s$^{-1}$ at $3.36\times10^{14}$ Hz.

These upper limits place some constraints on the proposed models for electromagnetic counterparts of BBH mergers. 
The prompt emission luminosity upper limit (from the BAT) is about one order of magnitude higher than the claimed luminosity of GW150914-GBM \citep{Connaughton2017}, so this upper limit is not very constraining for prompt GRB models for BBH mergers \citep{Loeb2016,Perna2016,Zhang2016}. 
Nevertheless, one may constrain the upper limit of the dimensionless charge $\hat q$ of the BHs. 
Using Equation (11) of \cite{Zhang2019}\footnote{We only consider the electric dipole radiation luminosity and neglect the magnetic dipole radiation luminosity (Equation (17) of \cite{Zhang2019}) to constrain $\hat q$. This is because the $a$-dependence is very steep for the latter component. Even if the instantaneous luminosities are comparable for the two components at the merger time, the total emitted energy from the latter is much smaller than the former. Therefore, it can be neglected when the average luminosity is considered for comparison with the luminosity upper limit from  observations.} and assuming that both BHs have roughly the same mass and $\hat q$ (noting $a=2 r_s$ at the merger, where $r_s$ is the Schwarzschild radius), one can derive $\hat q^2 ({c^5}/{G})/48 < 1.4 \times 10^{50} \ {\rm erg \ s^{-1}}$, which gives a shallow upper limit on the dimensionless charge
\begin{equation}
    \hat q < 1.4 \times 10^{-4}.
\end{equation}
This upper limit is higher than the value required to interpret GW150914-GBM, $\hat q \sim 5 \times 10^{-5}$, which has been regarded as extremely large for stellar mass black holes \citep{Zhang2016}. 
So this upper limit is not very constraining. 
The non-detection of bright gamma-rays at the merger time also disfavors bright emission from extreme scenarios, such as forming two black holes in a collapsing star \citep{Loeb2016} or reactivating a massive dead disk right after the merger \citep{Perna2016}, both requiring quite specific physical conditions. 

The X-ray upper limit can place an upper limit on the total energy of a putative electromagnetic explosion associated with the BBH merger event. 
Using Equation (20) of \cite{Wang2015}, with an X-ray flux upper limit of $4.2\times 10^{-12}$ erg cm$^{-2}$ s$^{-1}$ at 22 ks after the merger\footnote{The luminosity upper limit corresponds to an 80 s XRT exposure. 
Since the constraint on $E$ is tighter at earlier epochs, we adopt the earliest epoch, 22 ks (post-merger), in the calculation to reach the most stringent constraint on $E$.}, one can obtain an upper limit on isotropic-equivalent energy of the blast wave (see also \citealt{Perna2019})
\begin{equation}
    E < 4.1 \times 10^{51} \ {\rm erg} 
\end{equation}
assuming typical GRB parameters: electron energy distribution power-law index $p=2.2$, electron energy fraction $\epsilon_e = 0.1$, magnetic energy fraction $\epsilon_B = 0.01$, and circum-burst number density $n=1$ cm$^{-3}$ for the source\footnote{This estimate assumed $\nu_m < \nu < \nu_c$, which has a dependence of $n^{-2/(p+3)}$. For $\nu > \nu_c$, the upper limit is similar, but the result does not dependent on $n$.}. 
This rules out a bright GRB-like explosion, but a relatively faint explosion (e.g., a low-luminosity GRB) is not ruled out.
For comparison, the isotropic-equivalent energy released by GRB 170817A (GW 170817) was estimated to be $E\sim1\times10^{53}$ erg

Finally, the peak flux of the optical flare peaking $\sim 50$ d after GW 190521g (another BBH merger) is $\sim 10^{45} \ {\rm erg \ s^{-1}}$ \citep{Graham2020}.
The optical/near-UV upper limits we have derived here for S200224ca (from the Subaru/HSC coverage) are lower by at least 2 orders of magnitudes, even though the time of follow-up observations was much earlier than 50 d. 
This disfavors the AGN disk interaction model for an early interaction (assuming, of course, that the location of the putative optical counterpart to S200224ca was observed).

\section{SUMMARY}
During its third observing run, LVC detected S200224ca: a candidate GW event with a low false-alarm rate (roughly once every 1,974 years) and a low probability of being terrestrial ($P_{\rm Terrestrial} = 3.3952\times10^{-5}$).
This trigger was confidently identified to be produced by a BBH merger at an estimated distance $d=1575\pm322$ Mpc.
The event was very-well localized, with 50\% and 90\% probability areas corresponding to 17 and 69 deg$^2$, respectively.

At $T_0$, the {\sl Swift}-BAT FOV covered 88.38\% of the integrated GW localization region (with partial coding fraction $>$10\%).
No significant ($\gtrsim 5\sigma$) detections were seen in the BAT raw light curves.
Using the light curve with 1.6 s bins, we place a $5\sigma$ upper flux limit of $<4.76\times10^{-7}$ erg cm$^{-2}$ s$^{-1}$ (in 15--350 keV) on prompt emission from the merger within $T_0\pm100$ s.
No significant detections were seen in the survey data either (data binned to intervals of $\sim300$ s), between $T_0-115$ to $T_0-22$ s, and $T_0+180$ to $T_0+480$ s.
Looking forward, however, upon notice of GW triggers the newly-implemented BAT GUANO system \citep{Tohuvavohu2020} will continue to attempt to automatically retrieve event data around trigger times, allowing for much more sensitive searches for prompt (sGRB-like) emission from CBC events, and will increase the number of possible co-detections of GW and EM radiation.
Currently, the GUANO event recovery success rate has now reached $\sim90\%$.

{\sl Swift} performed targeted follow-up observations of the GW error region with the XRT and UVOT from $T_0+22$ ks to $T_0+196.5$ ks.
The XRT and UVOT observed 672 fields for approximately 80 s each.
The XRT covered 64.5 deg$^2$, corresponding to approximately 80\% of the galaxy-convolved (and 82\% of the raw) probability region, making S200224ca the BBH event most thoroughly followed-up in X-rays to date.

Although no likely EM counterparts were detected by {\sl Swift} BAT, XRT, or UVOT, nor by other facilities in any wavelength, these searches serve as observational evidence supporting the expected (null) result.
From the BAT data, we place a gamma-ray (15--350 keV) luminosity upper limit of $1.4_{-0.5}^{+0.6}\times10^{50}$ erg s$^{-1}$ on S200224ca at $T_0\pm100$ s.
From the XRT observations, we place X-ray (0.3--10 keV) luminosity upper limits of $1.2_{-0.4}^{+0.6}\times10^{45}$ and $3.0_{-1.1}^{+1.3} \times10^{45}$ erg s$^{-1}$ (at 50\% and 90\% confidence) on S200224ca from $T_0+22$ ks to $T_0+196.5$ ks.
From the UVOT $u$-band observations, we place a luminosity ($\nu L_\nu$) upper limit of $7.24\times10^{43}$ erg s$^{-1}$ (at $8.65\times10^{14}$ Hz).
We calculate tighter limits (albeit at redder wavelengths) from optical ground-based observations (e.g., Subaru-HSC; \citealt{Ohgami2020}), which place  luminosity $\nu L_\nu < 6.2^{+2.7}_{-2.2}\times10^{41}$ erg s$^{-1}$ at $4.82\times10^{14}$ Hz ($r$-band), and $<1.43^{+0.65}_{-0.52}\times10^{42}$ erg s$^{-1}$ at $3.36\times10^{14}$ Hz ($z$-band).

From these limits, we place a shallow upper limit on the dimensionless BH charge, $\hat{q}<1.4\times10^{-4}$, and an upper limit on the isotropic-equivalent energy of a blast wave from the merger $E<4.1\times10^{51}$ erg (assuming typical GRB parameters).
These limits also disfavor bright GRB-like emission from the BBH merger, but still allow for faint emission (e.g., a low-luminosity GRB).
Finally, the non-detection of gamma-rays disfavors some hypothetical exotic scenarios, such as the formation of two black holes in a collapsing star \citep{Loeb2016}, reactivating a massive dead disk right after the merger \citep{Perna2016}, or BHs with an extremely large charge in excess of $\hat q = 1.4 \times 10^{-4}$.

\facility{{\sl NASA Neil Gehrels Swift Observatory}}

\acknowledgements

This publication is an official product of the {\sl Swift} GW follow-up team.
The authors wish to thank the anonymous referee for their helpful comments which enhanced the clarify of the paper.
NJK and JAK would like to acknowledge support from NASA Grants 80NSSC19K0408 and 80NSSC20K1104.
PAE, KLP, APB, JPO, AAB, SWKE, NPMK, and MJP acknowledge funding from the UK Space Agency.
PDA acknowledge support from PRIN-MIUR 2017 (grant 20179ZF5KS).
MGB, GC, SC, ADA, PDA, AM and GT acknowledge funding from the Italian Space Agency, contract ASI/INAF n.\ I/004/11/4.
The authors would like to acknowledge support by from the Italian Ministry of Foreign Affairs and International Cooperation grant MAE0065741.
ET acknowledges support by the National Aeronautics and Space Administration through grant NNX10AF62G issued through the Astrophysics Data Analysis Program.
MDP acknowledges support for this work by the Scientific and Technological Research Council of Turkey (T\"UB\.ITAK), Grant No: MFAG-119F073.
DBM is supported by research grant 19054 from Villum Fonden.

\end{document}